\documentclass[a4paper,11pt]{article}
\usepackage{pos}
\usepackage{cleveref}
\usepackage{natbib}
\usepackage{xspace}
\usepackage{floatrow}

\newcommand{\mtL}{m_{\tilde{t}_L}}
\newcommand{\mtR}{m_{\tilde{t}_R}}

\newcommand{\SUSY}{{\text{SUSY}}}
\newcommand{\MSSM}{{\text{MSSM}}}

\newcommand{\msusy}{\ensuremath{M_\SUSY}\xspace}

\newcommand{\tev}{\;\text{TeV}\xspace}
\newcommand{\OS}{{\ensuremath{\text{OS}}}\xspace}

\newcommand{\DR}{{\ensuremath{\overline{\text{DR}}}}\xspace}
\newcommand{\MDR}{{\ensuremath{\overline{\text{MDR}}}}\xspace}
\newcommand{\fin}{{\ensuremath{\text{fin}}}\xspace}
\newcommand{\als}{\alpha_s}
\newcommand{\alt}{\alpha_t}
\newcommand{\xt}{\ensuremath{\widehat X_t}\xspace}

\title{Experimental probes and theoretical concepts for BSM trilinear couplings: a case study for scalar top quarks }
\ShortTitle{Experimental probes and theoretical concepts for BSM trilinear couplings}

\author[a,b]{Henning Bahl}
\author*[c]{Johannes Braathen}
\author[c,d]{Georg Weiglein}

\affiliation[a]{University of Chicago, Department of Physics, 5720 South Ellis Avenue, Chicago, IL~60637~USA}
\affiliation[b]{Institut für Theoretische Physik, Philosophenweg 16, 69120 Heidelberg, Germany}
\affiliation[c]{Deutsches Elektronen-Synchrotron DESY, Notkestr. 85, 22607 Hamburg, Germany}
\affiliation[d]{II. Institut für Theoretische Physik, Universität Hamburg, Luruper Chaussee 149, 22761 Hamburg, Germany}

\emailAdd{bahl@thphys.uni-heidelberg.de}
\emailAdd{johannes.braathen@desy.de}
\emailAdd{georg.weiglein@desy.de}

\abstract{
After the possible discovery of new particles, it will be crucial to determine the properties, and in particular the couplings, of the new states. Here, we focus on scalar trilinear couplings, employing as an example the case of the trilinear coupling of scalar top quarks (stops) to the Higgs boson in the Minimal Supersymmetric Standard Model (MSSM). We discuss possible strategies for experimentally determining the stop trilinear coupling parameter, which controls the stop--stop--Higgs interaction, and we demonstrate the impact of different prescriptions for the renormalisation of this parameter. We find that the best prospects for determining the stop trilinear coupling arise from its quantum effects entering the model prediction for the mass of the SM-like Higgs boson in comparison to the measured value, pointing out that the prediction for the Higgs-boson mass has a high sensitivity to the stop trilinear coupling even for heavy masses of the non-standard particles. Regarding the renormalisation of the stop trilinear coupling, we identify a renormalisation scheme that is preferred given the present level of accuracy, and we clarify the origin of potentially large logarithms that cannot be resummed with standard renormalisation group methods.
}

\FullConference{The European Physical Society Conference on High Energy Physics (EPS-HEP2023)\\
 21-25 August 2023\\
Hamburg, Germany\\}

\begin{document}

\begin{flushright}
DESY-23-164
\end{flushright}

\maketitle

\section{Introduction\vspace{-0.3cm}}
The Higgs boson at 125 GeV is so far the only known scalar particle without a known substructure. However, many extensions of the Standard Model (SM) introduce additional scalar degrees of freedom in order to address questions left unresolved in the SM, $e.g.$ the nature of dark matter, or the observed baryon asymmetry of the Universe. A particularly interesting new type of interaction potentially arising in Beyond-the-SM (BSM) models is a trilinear scalar interaction that is not generated by a vacuum expectation value but instead arises as a consequence of dimensionful couplings, independently of spontaneous symmetry breaking. Such dimensionful couplings appear for instance in supersymmetric (SUSY) theories --- like the Minimal Supersymmetric SM (MSSM) --- which predict trilinear couplings between the Higgs bosons and the SUSY partners of the SM fermions. Among these, the trilinear coupling between the stops (the SUSY partners of top quarks) and the SM-like Higgs boson is of particular importance. This ``stop mixing parameter'' is typically the largest of the trilinear couplings and it controls the Higgs--stop--stop interaction itself as well as the mass splitting between the stops.
 
If an extended scalar sector is discovered at the LHC or a future collider, the measurements of the interactions between the various scalars will be crucial to pinpoint the underlying theory. With this motivation, we discussed in Ref.~\cite{Bahl:2022kzs}, for the specific case of the MSSM stop mixing parameter $X_t$, how this coupling could be extracted from experimental measurements, and how to define it properly in theoretical predictions. We investigated different methods to extract the stop mixing parameters from experiments, pointing out difficulties of the various approaches and emphasising the crucial role of the mass of the SM-like Higgs boson. In turn, we examined appropriate choices of renormalisation schemes for the stop mixing parameter in Higgs-mass calculations. In this process, we clarified the origin of large Sudakov-like logarithms plaguing Higgs-mass predictions in the on-shell scheme when combining fixed-order and EFT techniques. To avoid their occurrence, we proposed to use a mixed scheme where the stop mixing parameter is renormalised in the \DR/\MDR scheme, while the stop masses are renormalised on-shell.

\vspace{-0.3cm}

\section{Experimental probes of $X_t$\vspace{-0.3cm}}
We begin by reviewing in this section different possible methods to access the stop mixing parameter $X_t$ experimentally. We refer the reader to $e.g.$ Ref.~\cite{Bahl:2022kzs} for a description of the stop sector in the MSSM. We recall that $X_t$ is defined as $X_t\equiv A_t-\mu \cot\beta$, with $A_t$ the soft SUSY-breaking stop trilinear coupling, $\mu$ the Higgsino mass parameter (assumed here to be real for simplicity) and $\tan\beta\equiv t_\beta$ the ratio of the vacuum expectation values of the two Higgs doublets. A first option to access $X_t$ is to consider the stop mass eigenvalues, which depend on $X_t$ as
\begin{align}
m_{\tilde t_{1,2}}^2 = m_t^2 + \frac{1}{2}\left\{m_{\tilde{t}_L}^2 + m_{\tilde{t}_R}^2 \mp \sqrt{\left[m_{\tilde{t}_L}^2 - m_{\tilde{t}_R}^2 + M_Z^2 c_{2\beta} \left(\frac{1}{2} - \frac{4}{3}s_W^2\right)\right]^2 + 4 m_t^2 |X_t|^2}\right\}.
\end{align}
Here $\mtL$ and $\mtR$ denote the soft SUSY-breaking stop mass parameters, $m_t$ and $M_Z$ the top-quark and $Z$-boson masses respectively, and $s_W$ is the sine of the weak mixing angle. 
The dependence of these mass eigenvalues on $X_t$ is, however, significantly reduced as $\msusy\equiv \sqrt{\mtL \mtR}$ increases. Additionally, extracting $X_t$ from $m_{\tilde{t}_{1,2}}$ requires some extra information or assumption about the relation between the soft stop mass parameters: indeed, the knowledge of two mass eigenvalues is obviously not enough to extract the three parameters $\mtL$, $\mtR$, and $X_t$. Supposing that one were to obtain a third input in the form of the stop mixing angle $\theta_{\tilde{t}}$ (by itself a challenging quantity to measure~\cite{Rolbiecki:2009hk}), one would only find sensitivity to $X_t$ if there is a significant splitting between $\mtL$ and $\mtR$. Moreover, the dependence of $\theta_{\tilde{t}}$ on $X_t$ becomes very small for increasing $\msusy$. In other words, for high $\msusy$, even if one could build a collider allowing to measure the stop masses $m_{\tilde{t}_i}$ and the mixing angle $\theta_{\tilde{t}}$, it would still not be possible to obtain accurate information about $X_t$. 

\begin{figure}
\centering
\begin{minipage}{.42\textwidth}
\includegraphics[width=\textwidth]{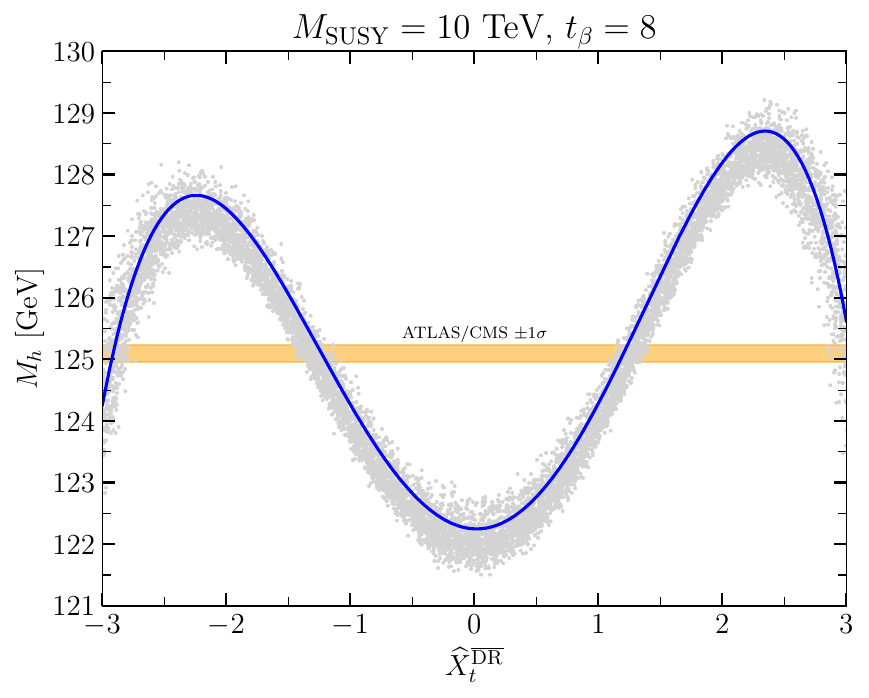}
\end{minipage}
\begin{minipage}{.42\textwidth}
\includegraphics[width=\textwidth]{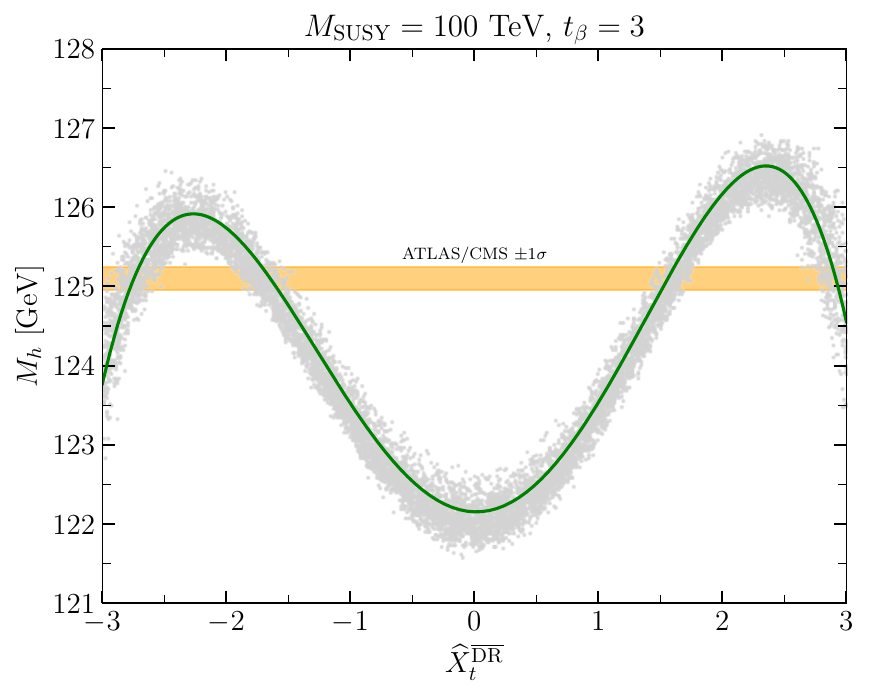}\vspace{-.2cm}
\end{minipage}
\caption{$M_h$ as a function of $\widehat X_t\equiv X_t/\msusy$. \textit{Left}: The blue curve displays $M_h$ for $t_\beta=8$ in a single-scale scenario with $\msusy=10\tev$, where all BSM mass terms are set to $\msusy$ and all trilinear couplings, except $A_t$, are set to zero. The gray points are obtained by varying mass parameters and trilinear couplings $A_{f\neq t}$ randomly within $[1/2 \msusy, 2\msusy]$. 
\textit{Right}: Same as left plot, but with $t_\beta=3$ and $\msusy = 100\tev$.\vspace{-0.5cm}}
\label{fig:Mh}
\end{figure}

Another option is to consider decay processes in which the parameter $X_t$ enters at the tree level, such as $\tilde{t}_2\to\tilde{t}_1 h$. Such a decay process is, however, only useful if both its corresponding decay width and branching ratio are sufficiently large to allow precise measurements, 
which depends strongly on the SUSY mass spectrum. For instance, if the stops are too close in mass, or if other decay channels are open ($e.g.$ if there are relatively light electroweakinos), then the $\tilde{t}_2\to\tilde{t}_1 h$ decay does not allow a reliable extraction of $X_t$ --- as was also illustrated numerically in Ref.~\cite{Bahl:2022kzs}.

Finally, while it may seem like $X_t$ simply has no further phenomenological impact if $\msusy$ becomes large, there is one observable that does exhibit a significant dependence on $X_t$ even for large $\msusy$: the Higgs-boson mass $M_h$. Indeed in SUSY theories, because of the additional underlying symmetry, $M_h$ can be computed as a function of model parameters --- see $e.g.$ the recent review~\cite{Slavich:2020zjv}. Together with $\msusy$ and $t_\beta$, the stop mixing parameter $X_t$, which enters the prediction of $M_h$ from one-loop order, is among the quantities with the strongest influence on $M_h$. This is illustrated in \cref{fig:Mh} where $M_h$, computed with \texttt{FeynHiggs 2.18.1}~\cite{FeynHiggs,Hahn:2013ria}, is shown as a function of $\widehat X_t\equiv X_t/\msusy$, in scenarios with $\msusy=10\tev$ and $t_\beta=8$ (left) or $\msusy=100\tev$ and $t_\beta=3$ (right), and compared with the present $1\sigma$ experimental uncertainty (orange band). The blue (left) and green (right) curves correspond to the Higgs-boson mass in single-scale scenarios where all BSM masses are equal to $\msusy$ while all soft trilinear terms other than $A_t$ are set to zero. These results indicate that a comparison of the already very precise measurement of $M_h$ with high-precision predictions for it offers the best prospects to access $X_t$ experimentally. Moreover, the gray points, obtained by randomly scanning on masses and trilinear couplings (other than $A_t$) in the range $[1/2\msusy, 2\msusy]$, show that an imprecise knowledge of the SUSY spectrum would not severely degrade the extraction of $X_t$ from $M_h$. 
\vspace{-0.3cm}

\section{Renormalisation of $X_t$ in Higgs-boson mass calculations\vspace{-0.3cm}}
Having shown that the Higgs-boson mass offers the best possibility to access $X_t$, we investigate now appropriate choices of renormalisation schemes for $X_t$ in the context of Higgs-mass computations. Detailed discussions of renormalisation schemes for $X_t$ can be found $e.g.$ in Refs.~\cite{stopren}, while the different approaches for Higgs-mass calculations are reviewed at length in Ref.~\cite{Slavich:2020zjv}. 
Three main methods can be distinguished: \textit{fixed-order calculations}, reliable for low \msusy and in which a full on-shell (OS) renormalisation is possible; \textit{EFT calculations}, allowing a consistent resummation of large logarithms occurring for higher values of \msusy --- EFT computations are, however, usually performed only to leading order in the $v/\msusy$ expansion (see Ref.~\cite{Bagnaschi:2017xid} for an exception) and are therefore less reliable for low \msusy; and lastly, \textit{hybrid calculations}, combining the fixed-order approach with an EFT-like resummation of large logarithms. For EFT computations, a \DR (or an \MDR\footnote{The \MDR scheme~\cite{mdrscheme} redefines the finite parts of the stop sector counterterms in order to avoid unphysical non-decoupling effects if the gluino is significantly heavier than the stops. }) scheme is preferred to prevent large logarithms in the calculation of threshold corrections. In hybrid calculations, an OS renormalisation is simple to implement in the fixed-order part of the calculation, while a \DR or \MDR one is preferred in the EFT part. This shows the need of a renormalisation-scheme conversion of $X_t$ in typical hybrid calculations. 

\begin{figure}
\floatbox[{\capbeside\thisfloatsetup{capbesideposition={right,center},capbesidewidth=4cm}}]{figure}[\FBwidth]
{\caption{Comparison of $\delta^{(1)} (m_t X_t)/ (m_t \mtL)$ calculated without any expansion (red solid line), calculated in the limit $v/\msusy\to 0$ with $\mtL\neq\mtR$ (blue dashed), and calculated in the limit $v/\msusy\to0$ with $\mtL = \mtR$ (black dotted).}\label{fig:dXtmt}}
{ \includegraphics[width=.55\textwidth]{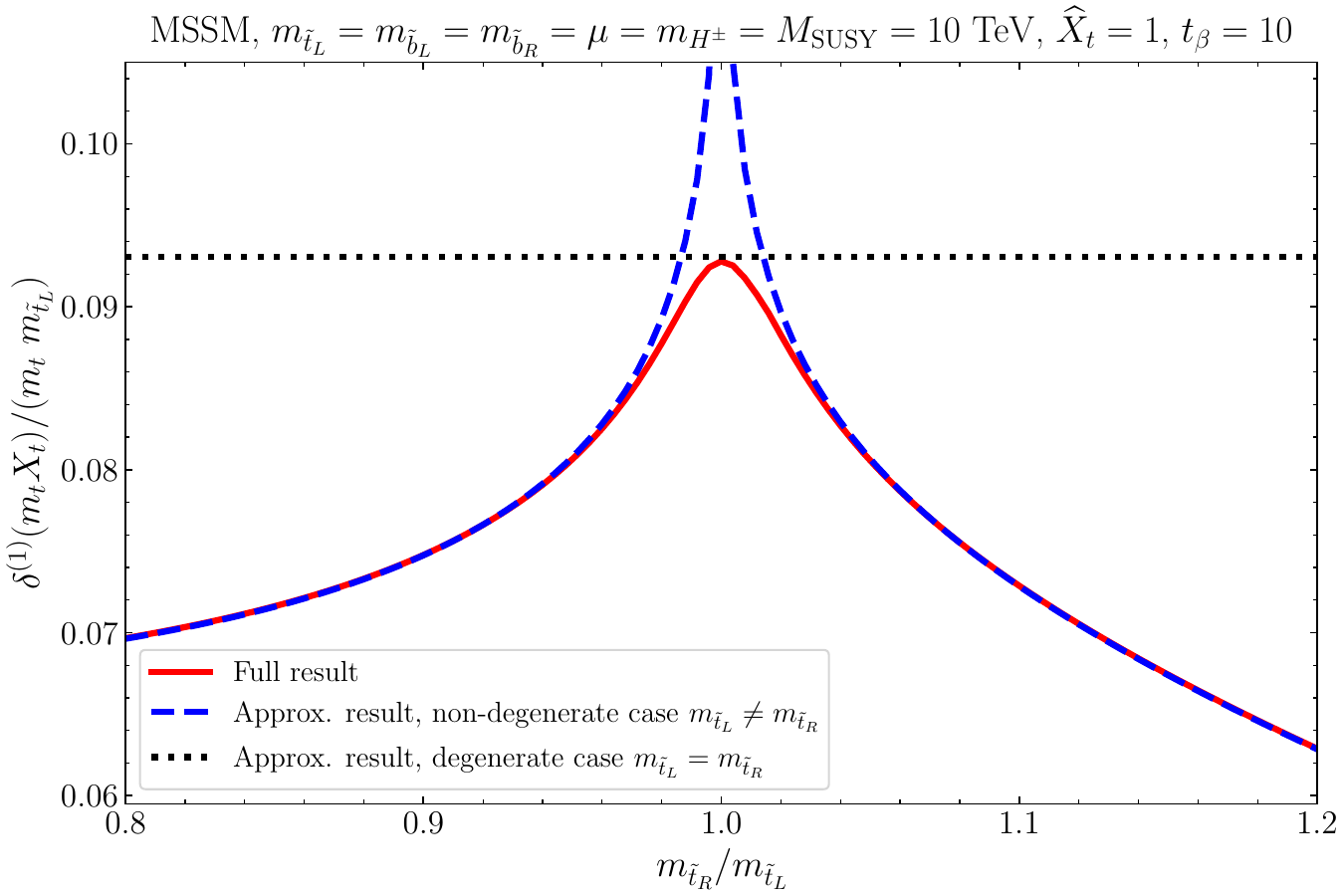}} 
\end{figure}

This conversion is, however, plagued by large logarithms: indeed, the relation between $X_t$ in the OS and in the \DR schemes is\vspace{-0.2cm}
\begin{align}
\label{eq:Xt_conv}
X_t^\OS =  X_t^{\DR}(\msusy) \ \frac{m_t^{\DR,\MSSM}(\msusy)}{m_t^\OS} - \frac{1}{m_t^\OS} \delta^{(1)} (m_t X_t) \bigg|_\fin.\vspace{-0.2cm}
\end{align}
Both terms on the right-hand side of this equation contain large logarithms, and while for the first term the logarithms can be resummed via renormalisation-group running, this is not the case for the second term. This term indeed is more complicated, in particular as it is necessary to expand it in powers of $v/\msusy$ in the context of an EFT or hybrid calculation --- in order not to mix orders in the EFT expansion. Depending on whether one considers $\mtL$ and $\mtR$ equal or not, one obtains two different expressions for the $\mathcal{O}(\alt)$ corrections to $\delta^{(1)}(m_tX_t)$, namely\vspace{-.3cm}
\begin{subequations}
\begin{align}
    \left.\delta^{(1)} (m_t X_t) \right|_\fin& \overset{\mtL=\mtR}{=} \frac{3\alt}{16 \pi} \ m_t X_t \; \vert \widehat{X}_t \vert^2 \ln \frac{\msusy^2}{m_t^2} + \cdots, \label{eq:Xt_conv_deg_logs}\\
    \left.\delta^{(1)} (m_t X_t) \right|_\fin& \overset{\mtL\neq \mtR}{=} {}\frac{\alt}{8 \pi} \; m_t X_t \ \vert \xt \vert^2 \; \left(\frac{2\mtL}{\mtR} \ln \frac{\mtL^2}{\vert \mtL^2 - \mtR^2 \vert} +\frac{\mtR}{\mtL} \ln \frac{\mtR^2}{\vert \mtL^2 - \mtR^2 \vert}  \right) + \cdots. \label{eq:Xt_conv_nondeg_logs}\vspace{-.3cm}
\end{align}
\end{subequations}
In both lines, the ellipsis denotes terms without large logarithms. Importantly, one finds that \cref{eq:Xt_conv_deg_logs} is not recovered when taking the limit $\mtR\to\mtL$ in \cref{eq:Xt_conv_nondeg_logs}.\footnote{We note that this behaviour differs from that of the $\mathcal{O}(\als)$ corrections, involving only diagrams with fermions or gauge bosons, in which a smooth limit exists.} This discrepancy in the behaviour of the two results is also illustrated in \cref{fig:dXtmt}, which shows $\delta^{(1)}(m_tX_t)/(m_t\mtL)$ as a function of the ratio $\mtR/\mtL$. The black dotted line corresponds to the result of \cref{eq:Xt_conv_deg_logs}, the blue dashed line to \cref{eq:Xt_conv_nondeg_logs}, while the red solid line is the full result (without an EFT expansion in $v/\msusy$). The large logarithms in \cref{eq:Xt_conv_deg_logs,eq:Xt_conv_nondeg_logs} were found in Ref.~\cite{Bahl:2021rts} to be connected to infrared (IR) singularities stemming from external-leg corrections --- and in this language, \cref{eq:Xt_conv_deg_logs,eq:Xt_conv_nondeg_logs} differ by the way in which the IR divergences are regulated (see Ref.~\cite{Bahl:2021rts}). While it is expected that a resummation of these large logarithms can be achieved with soft-collinear effective field theory, an explicit two-loop calculation in Ref.~\cite{Bahl:2021rts} has shown that logarithmic corrections of this type are typically of moderate size beyond one loop. Nevertheless, the presence of sizeable terms, as well as the lack of a smooth transition between the $\mtL=\mtR$ and $\mtL\neq\mtR$ cases, indicate that avoiding the need for a scheme conversion within hybrid calculations would be desirable. Therefore, our conclusion is that it is preferable to employ a \DR/\MDR renormalisation scheme for $X_t$ in both fixed-order and EFT parts of hybrid computations of $M_h$. We note, however, that mixed \DR-OS schemes can give rise to complications from uncancelled $\epsilon^1$ pieces of loop integrals at higher orders --- see Ref.~\cite{Bahl:2022kzs}
\vspace{-0.5cm}

\section{Summary\vspace{-0.3cm}}
An interesting possibility in models with extended scalar sectors is that of new types of interactions, like trilinear couplings that are not induced by a vacuum expectation value. We have focused here on the stop mixing parameter $X_t$ of the MSSM, which controls the interaction between the stops and Higgs bosons. We first discussed different approaches to access $X_t$ via experimental measurements, finding that $M_h$ offers the best way to determine $X_t$ (once the stop masses and $\tan\beta$ are known). Indeed, this observable exhibits sensitivity to $X_t$ irrespective of the stop mass hierarchy, and even for high $\msusy$. Decays of stops provide another option, but which is highly dependent on the sparticle spectrum ($i.e.$ which decay channels are open), and is thus only useful for parts of the parameter space. Next, we considered which choice of renormalisation scheme is most suitable for $X_t$, given the current state-of-the-art of Higgs-mass calculations. In particular, we investigated the conversion of $X_t$ from an OS scheme (convenient for fixed-order calculations) to the \DR/\MDR scheme (preferred for EFT calculations or the EFT part of hybrid calculations), and we clarified the origin of large logarithms appearing therein. We identified a class of large logarithms that cannot be resummed via standard EFT techniques and whose form depends on the limit in which the EFT expansion is performed. This led us to conclude that for the determination of $X_t$ by confronting a hybrid calculation of $M_h$ with its measured value, it is most advantageous to adopt, in both fixed-order and EFT parts of the computation, a \DR/\MDR renormalisation for $X_t$. 

\vspace{-.2cm}
\paragraph*{Acknowledgements}
\sloppy{H.B.\ acknowledges support by the Alexander von Humboldt foundation. J.B.\ and G.W.\ acknowledge support by the Deutsche Forschungsgemeinschaft (DFG, German Research Foundation) under Germany‘s Excellence Strategy --- EXC 2121 ``Quantum Universe'' --- 390833306. J.B.\ is supported by the DFG Emmy Noether Grant No.\ BR 6995/1-1. This work has been partially funded by the Deutsche Forschungsgemeinschaft 
(DFG, German Research Foundation) --- 491245950. }


\end{document}